\documentclass[aps,twocolumn,superscriptaddress,amsmath,nofootinbib,longbibliography,amssymb,prx,10pt]{revtex4-2}
\usepackage{graphicx}% Include figure files
\usepackage{float}
\usepackage[normalem]{ulem}
\usepackage{dcolumn}% Align table columns on decimal point
\usepackage{bm}% bold math
\usepackage{physics}
\usepackage{wasysym}
\usepackage{comment}
\usepackage{placeins} 
\usepackage{xcolor, colortbl}
\definecolor{darkblue}{rgb}{0,0,0.6}
\newcolumntype{g}{>{\columncolor{lightgray}\centering\arraybackslash}p{0.8cm}}
\newcolumntype{w}{>{\centering\arraybackslash}p{0.8cm}}
\usepackage[colorlinks,linkcolor=darkblue,citecolor=darkblue,urlcolor=darkblue]{hyperref}

\newcommand{\montpellier}{Laboratoire Charles Coulomb (L2C), Universit\'e de Montpellier, CNRS, 34095 Montpellier, France}
\newcommand{\grenoble}{Univ. Grenoble Alpes, CNRS, LIPhy, 38000 Grenoble, France}
\newcommand{\gulliver}{Gulliver, CNRS UMR 7083, ESPCI Paris, PSL Research University, 75005 Paris, France}
\newcommand{\rev}[1]{{\color{black}#1}}
\newcommand{\revv}[1]{{\color{black}#1}}

\begin{document}

\title{Molecular motion at the experimental glass transition}

\author{Romain Simon}

\affiliation{\montpellier}

\author{Jean-Louis Barrat}

\affiliation{\grenoble}

\author{Ludovic Berthier}

\affiliation{\gulliver}

\date{\today}

\begin{abstract} 
We propose a novel \rev{computational} strategy to study the glass transition of molecular fluids. Our approach combines the construction of simple yet realistic models with the development of Monte Carlo algorithms to accelerate equilibration and sampling. Inspired by the well-studied ortho-terphenyl glass-former, we construct a molecular model with an analogous triangular geometry and construct a `flip' Monte Carlo algorithm. We demonstrate that the flip Monte Carlo algorithm achieves a sampling speedup of about $10^9$ at the experimental glass transition temperature $T_g$. This allows us to systematically analyze the equilibrium structure and molecular dynamics of the model over a temperature regime previously inaccessible. We carefully compare the observed physical behavior to earlier studies that used atomistic models. In particular, we find that  the glass fragility and the departure from the Stokes-Einstein relation are much closer to experimental observations. We characterize the development and temperature evolution of spatial correlations in the relaxation dynamics, using both orientational and translational degrees of freedom. Excess wings emerge at intermediate frequencies in dynamic rotational spectra, and we directly visualize the corresponding molecular motion near $T_g$. Our approach can be generalized to a \rev{broad range of molecular geometries and paves the way to a deeper} understanding of how molecular details may affect more universal physical aspects characterizing molecular liquids approaching their glass transition.
\end{abstract}

\maketitle

\section{Introduction}

The physics of the glass transition, observed in different materials, remains a central research problem~\cite{ediger1996supercooled,debenedetti2001supercooled,berthier2016facets}, with constant progress on theoretical~\cite{Berthier2011,parisi2020theory}, computational~\cite{liu2022challenges,berthier2023modern} and experimental~\cite{hunter2012physics,Ediger2017highly} fronts. Experiments mostly study molecular fluids, access a broad range of timescales from picoseconds to several hours, but cannot easily resolve the dynamics of molecules in space and time near the experimental glass transition temperature $T_g$, where the relaxation dynamics is very slow . By contrast, computer simulations have largely (but not exclusively) focused on atomistic models, the motion of atoms is by construction easily resolved in space and time, but conventional numerical techniques (mostly molecular dynamics) are limited to short timescales and fall out of equilibrium at temperatures much higher than the experimental $T_g$. 

To close the gap between simulations and experiments, one should analyze models that resemble more closely the molecular systems studied experimentally, but also improve the simulation techniques themselves to be able to access temperatures closer to $T_g$ and longer timescales. Progress is constantly being made to improve the simulated models~\cite{liu2022challenges}. Multicomponent and continuously polydisperse atomistic models may be seen as meaningful models for metallic glasses and colloidal particles, respectively. Molecules, on the other hand, can be simulated using either coarse-grained models~\cite{lewis1994molecular,sciortino1996supercooled,michele2001viscous,mazza2006relation,lombardo2006computational,chong2009coupling,fragia2015rotational}, or all-atom representations with accurate force fields~\cite{Eastwood2013,henot2023orientational}. The increasing complexity of these models, compared to point particles, comes with an additional computational cost. \rev{Despite decades of research, no universal computational method exists that can equilibrate any liquid (even defined by coarse-grained interactions) approaching its glass transition. Conventional molecular dynamics remains to date the only broadly applicable technique, but its severe timescale limitations are well known~\cite{allen2017computer}.} 

Regarding simulation techniques, hardware improvement or novel architectures have provided some acceleration~\cite{bailey2017rumd}. A remarkable achievement, relevant to our work, is the use of the massively parallel Anton computer to simulate an all-atom model for ortho-terphenyl, where simulated timescales near 1$\mu$s could be accessed~\cite{Eastwood2013}. A second path is the development of efficient algorithms that can speedup the equilibration of deeply supercooled states, and many ideas are currently being explored with a range of accomplishments~\cite{barrat2023computer,berthier2024monte,Jung_2024}. These algorithms have been however mostly applied to atomistic models. \rev{So far, the only algorithm able to reach experimental conditions for some (but not all) atomic models is the swap Monte Carlo algorithm~\cite{Grigera2001} which achieves an extreme speedup for a range of tailored models~\cite{berthier2016equilibrium,ninarello2017models,parmar2020ultrastable,jung2023predicting,alvarez2023simulated}. Using Swap Monte Carlo, equilibration near the experimental glass transition can be achieved, allowing direct comparisons to experiments~\cite{Berthier2017, Ozawa2018, Guiselin2022}.} 

An important task is thus to extend progress regarding algorithms to the realm of molecular fluids, \rev{but progress in this direction has been slow. Ozawa {\it et al.}~\cite{Ozawa2023,Ozawa2025} introduced a `random bonding' approach to create stable glassy configurations of dimers, however with a broad and uncontrolled bond length dispersity. B\"ohmer {\it et al.}~\cite{bohmer2025swap} introduced a size polydisperse model of dumbbells and performed swap moves where pairs of dumbbells with unequal sizes are exchanged. These authors admit that the efficiency will ``decrease with increasing complexity of the molecular models"~\cite{bohmer2025on}. In addition, in both approaches, the produced configurations contain an unwanted source of disorder (bond lengths, dumbbell sizes), whereas molecular fluids are instead composed of identical molecules. These encouraging approaches thus remain fairly limited in scope.}  

\rev{By contrast, here we introduce a generic strategy similar in performance to swap Monte Carlo, but now applied to coarse-grained models of molecular fluids composed of identical molecules, and for which swap Monte Carlo cannot be applied. Following this approach, we propose that a broad class of molecular glass models can now potentially be equilibrated using a flexible Monte Carlo strategy with a speedup comparable to that of the swap Monte Carlo, for a broad range of molecular sizes or architectures, up to macromolecular polymer glasses. In this report, we focus on a specific example inspired by ortho-terphenyl, a molecular glass-former that has been the topic of numerous experimental investigations.}

Why is this endeavor an important research goal? There are several scientific reasons to justify its relevance. A first justification lies in the nature of the measurements most commonly performed in experiments. Experimental relaxation spectra are dominated by dielectric techniques that are mostly sensitive to rotational degrees of freedom rather than translational ones~\cite{kudlik1999dielectric,lunkenheimer2000glassy}. A second, related point, is the extent of dynamic decoupling between different degrees of freedom, which questions how rotations and translations are coupled as in the Stokes-Einstein-Debye (SED) relation. Decoupling and violations of the SED relation have been debated in particular in the context of dynamic heterogeneity~\cite{fujara1992translational,stillinger1994translation,cicerone1996enhanced,tarjus1995breakdown,ediger2000spatially}. It is in particular not established whether the multi-component nature of atomistic models plays a role in numerical results, compared to the  single component molecular fluids which are, by construction, structurally more homogeneous. A third issue with atomistic models is the gap between glass fragilities~\cite{angell1995formation} achieved numerically and experimentally,  with real molecular fluids showing much stronger deviations from Arrhenius slowing down compared to monoatomic models~\cite{sastry2001relationship,tarjus2004disentangling,alba2022}.

To tackle these issues, we propose a novel strategy to efficiently study numerical models for molecular fluids near the experimental glass transition. Regarding models, our work is deeply rooted in the idea of formulating coarse-grained molecular models, in the spirit of bead-spring models for polymer glasses~\cite{grest1986molecular} or simple molecules~\cite{lewis1994molecular}. \rev{Although the approach we propose remains for now limited to coarse-grained models, there are many open physics questions that can be attacked with such models as mentioned above, and emphasized again in the discussion in Sec.~\ref{sec:discussion}.} Regarding algorithms, we build on the recent success of the swap Monte Carlo algorithm~\cite{ninarello2017models}. We exemplify our approach on a specific model, which is directly inspired by existing coarse-grained models for ortho-terphenyl~\cite{lewis1994molecular}. We show that a small \rev{yet non-trivial} modification of this description produces a model composed of identical molecules with excellent glass-forming ability, and for which a `flip' Monte Carlo algorithm can be devised which achieves an efficiency comparable to that of the swap Monte Carlo. 

This strategy allows us to equilibrate large bulk systems even at temperatures below the experimental glass transition $T_g$, a regime previously inaccessible for molecular systems. Using these configurations, we characterize the structural properties at very low temperatures. Taking them as initial conditions, we can probe and resolve the motion of molecules over a broad dynamic range, up to the millisecond timescale. In particular, we propose an in-depth analysis of  the glass fragility, of deviations from the Stokes-Einstein relation, of dynamic heterogeneity and relaxation spectra, and propose quantitative comparisons to experimental results.

The manuscript is organized as follows. 
In Sec.~\ref{sec:model} we present the numerical model and the Monte Carlo algorithm developed in this work.  
In Sec.~\ref{sec:equilibration} we provide equilibrium static and dynamic correlation functions, and an estimate of the equilibration speedup of the flip MC algorithm. 
In Sec.~\ref{sec:dynamics} we characterize the fragility of the model, and discuss deviations from the Stokes-Einstein-Debye relation.
We characterize dynamic heterogeneity in Sec.~\ref{sec:dynamic}.
In Sec.~\ref{sec:excess} we discuss and interpret the emergence of excess wings in relaxation spectra. We conclude the paper in Sec.~\ref{sec:discussion} and offer perspectives for future work. 

\section{Molecular model and Monte Carlo algorithm}

\label{sec:model}

\subsection{Molecular model}

\label{sec:molecular}

\begin{figure}
\includegraphics[width=\columnwidth]{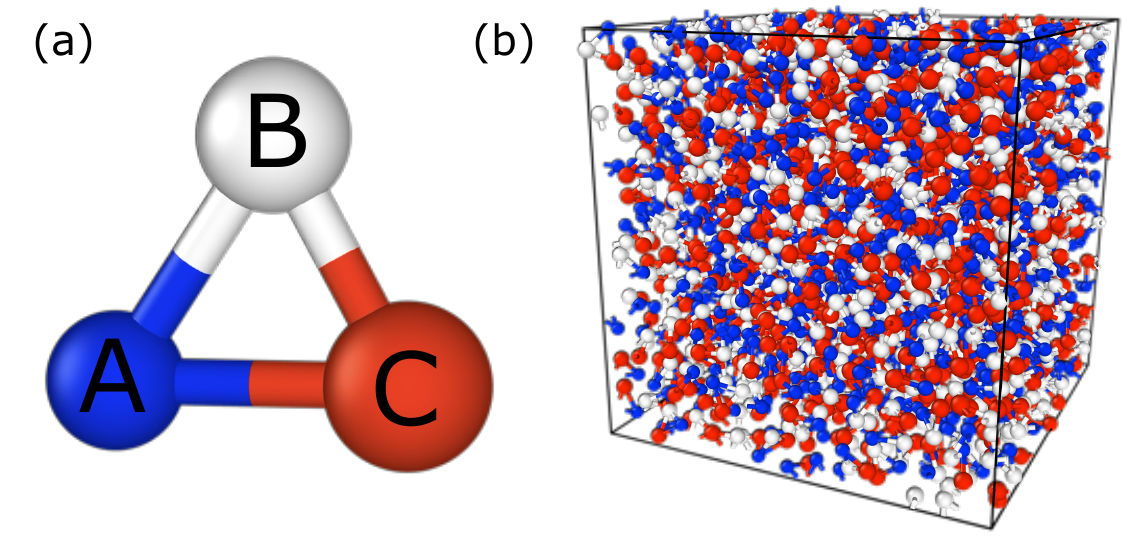}
\caption{(a) Schematic representation of the studied molecule with atoms $A$, $B$ and $C$ connected by non-rigid bonds. (b) Representative snapshot of the bulk simulated molecular fluid with $N=4000$ molecules.}
\label{fig:model}
\end{figure}

We construct a model composed of $M$ molecules, where each molecule is composed of three atoms $A$, $B$, $C$ of equal mass $m$ and diameters $\sigma_A$, $\sigma_B$ and $\sigma_C$, connected by bonds, as shown in Fig.~\ref{fig:model}(a). 

The interactions between atoms (both intramolecular and intermolecular) are governed by a purely repulsive Weeks-Chandler-Andersen (WCA) potential:
\begin{equation}
U_{\rm WCA}(r_{ij}) =
      4 \epsilon \Bigg(\Big(\frac{\sigma_{ij}}{r_{ij}}\Big)^{12} - \Big(\frac{\sigma_{ij}}{r_{ij}}\Big)^{6} + \frac{1}{4}\Bigg). 
\end{equation}
This potential applies when \( r_{ij} < 2^{1/6} \sigma_{ij} \), where \( r_{ij} \) is the distance between particles \( i \) and \( j \), and $\sigma_{ij}=\frac{1}{2}(\sigma_i + \sigma_j)$ is the additive diameter governing pair interactions. The potential vanishes at distances larger than this cutoff. 

To form bonds between atoms within the same molecule, we use an attractive finitely extensible nonlinear elastic (FENE) potential:
\begin{equation}
U_{\rm FENE}(r_{ij}) = -33.75 \epsilon \ln \Bigg(1 - \Big(\frac{r_{ij}}{1.5 \sigma_{ij}}\Big)^2\Bigg) .
\end{equation}
These FENE bonds act as non-linear springs which cannot be broken and cannot extend beyond $1.5 \sigma_{ij}$, as frequently used in bead-spring models for polymer chains~\cite{grest1986molecular}. This model resembles the model proposed by Lewis and Wahnstr\"om~\cite{lewis1994molecular}, with the difference that our triangular molecule is not fully rigid. We did not attempt to match the flexibility of the molecule to experimental measurements~\cite{kolbel2024ortho}. \revv{The chosen parameters are close to the ones conventionally used in studies of polymer glasses that also often use bead-spring models to represent chains~\cite{grest1986molecular,barrat2010molecular}. While we did not explore this path, these parameters could be further optimized to either improve the physical properties of the model, or the efficiency of the flip Monte Carlo algorithm.}

We set parameters and units as follows. We use $\sigma_B$ as the unit length, and define $\sigma_A=0.9$ and $\sigma_C=1.1$ so that the three atoms within the molecules have slightly different sizes, but all molecules are identical. We use $\epsilon$ as the unit of energy, and when using molecular dynamics (MD) simulations we \rev{express} times in units of $\sqrt{m \sigma_B^2 / \epsilon}$. \rev{Reduced units are used throughout the manuscript.}

We study three-dimensional systems composed of $N = 3000$ particles or, equivalently, $M = 1000$ molecules within a cubic simulation box of linear size $L$ so that the number density is $\rho = N/V = 1.2$. Periodic boundary conditions are applied. A representative snapshot of the  simulated system is shown in Fig.~\ref{fig:model}(b). To illustrate our results using snapshots, we also performed a few simulations using $N=4000$ molecules.  

The data analyzed in this work are available in a public repository~\cite{SimonZenodo2026}.

\subsection{Flip Monte Carlo algorithm}

Conventional Monte Carlo (MC) simulations for off-lattice fluids construct a Markov chain where simple trial moves are proposed, and are then accepted or rejected with a probability that is carefully constructed such that the stochastic process converges to the desired equilibrium Boltzmann distribution~\cite{frenkel2023understanding}. This is usually done by imposing the condition of detailed balance. Monte Carlo simulations for fluids typically propose a succession of small displacements of the atoms, leading to a dynamics that is physically very close to Brownian dynamics, and yields results that are indeed close to the ones obtained using molecular dynamics~\cite{berthier2007monte,berthier2024monte}.   

The decisive advantage of MC simulations is that ``non-physical" particle moves can be proposed, in the sense that the corresponding trajectories  would likely not be observed in MD simulations. Provided detailed balance is satisfied, the process converges to the same Boltzmann distribution, but the convergence to equilibrium and the exploration of configuration space can be enhanced if moves are well-chosen. This strategy is at the core of the swap MC algorithm where pairs of distinct atoms within the liquid can swap positions~\cite{Grigera2001}. Swap moves can be applied to our molecules, but since they are all identical this would provide no benefit, whereas swapping atoms would change the chemical composition of the molecules, and thus the nature of the studied system. The idea of constructing polydisperse molecules was recently explored~\cite{bohmer2025swap} but this then artificially reintroduces some degree of polydispersity between the relaxing objects. 

\begin{figure}
\includegraphics[width=\columnwidth]{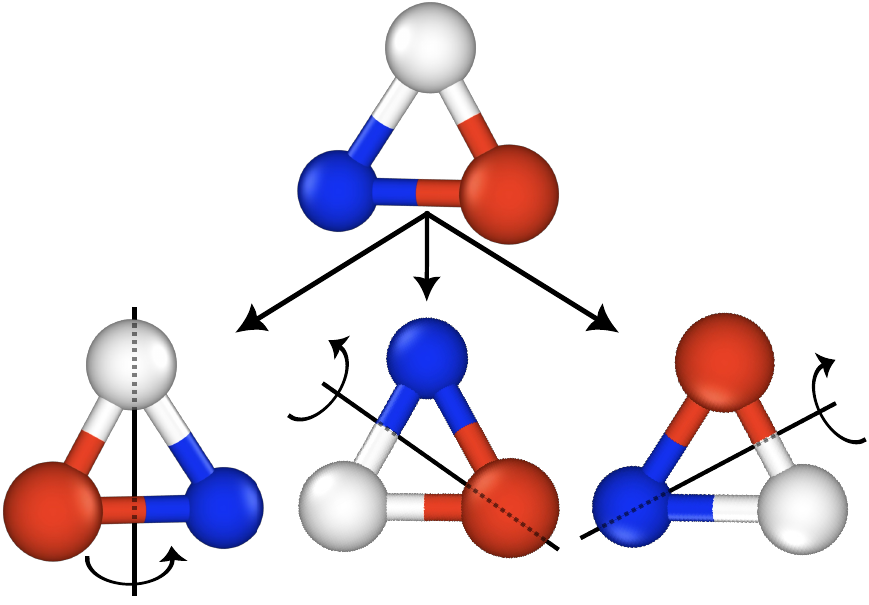}
\caption{The three Monte Carlo moves of the flip Monte Carlo algorithm. For a selected molecule three flips can be proposed, corresponding approximately to $180^\circ$ rotations around one the three axis shown with lines. Each flip corresponds to an intramolecular swap of two atoms exchanging their identity.}  
\label{fig:flip}
\end{figure}

Here, we introduce a different type of Monte Carlo moves that exploits the geometry of the molecule, see Fig.~\ref{fig:flip}. First we randomly select a molecule. We then randomly select one of the three median axis shown in Fig.~\ref{fig:flip}, that join one summit to the center of mass of the molecule. We then attempt to exchange the chemical nature of the two remaining summits. The flip move is accepted with the Metropolis acceptance rule, and detailed balance is therefore satisfied. Another way to describe such flip is to realize that it corresponds to an intramolecular swap Monte Carlo move where two distinct atoms of the molecule exchange diameters. We describe this as a `flip' move, as for an equilateral triangle it would exactly correspond to an 180 degrees rotation around the chosen axis. Note that as the atoms and bonds are slightly different, our molecules are not rigorously equilateral triangles: the flip move leaves the chemical nature of the system intact, but our working hypothesis is that, thanks to the slight asymmetry in shape, it could provide a speedup similar to the one demonstrated by swap MC for atomistic systems. Because our molecular fluid is composed of identical molecules, the need to introduce size disperse relaxing entities in swap MC simulations is relieved. 

We are now in a position to construct a flip Monte Carlo simulation scheme for the model described in Sec.~\ref{sec:molecular}. We combine flip moves, attempted with probability $p_{\text{flip}} = 0.2$ to conventional translational moves, attempted with probability $1-p_{\text{flip}} = 0.8$. Just as for swap MC, \rev{optimizing for a precise value of $p_{\text{flip}}$ is not very crucial~\cite{ninarello2017models}, provided it is neither too small (no flips) or too large (no translations).} 

\rev{There are two additional factors that control the efficiency of the flip Monte Carlo algorithm. Firstly, as for swap moves~\cite{ninarello2017models},  one must choose a size disparity for atoms within the molecules that results from a compromise. If it is too large, atom exchanges are never accepted, but if they are too small the accepted exchanges do not speedup the equilibration much. The  choice presented here turned out to be very efficient, hence  we have not specifically optimized this parameter. Secondly, thinking of molecular architectures with various numbers of atoms, we believe that the speedup will result from the competition between the number of degrees of freedom and the number of possible swap moves that leave the molecule unchanged. For the model under study, we have three atoms and three possible swap moves. Using a molecule with ABB atoms, for instance, would reduce the number of possible swap moves and presumably the algorithmic speedup. These considerations will guide the construction of future molecular models.}

We define a Monte Carlo timestep as $N$ attempted Monte Carlo moves (flip or translations). In practice, we use flip MC to carefully equilibrate the system at the desired temperature $T$. We then perform flip MC simulations to produce a large number of equilibrated configurations. In these simulations, the number of particles and the volume are kept constant, and we therefore explore the $NVT$ canonical ensemble. 

\subsection{Molecular dynamics simulations}

\label{sec:MD}

To study the physical dynamics of the system, we use equilibrium configurations produced using the flip MC algorithm as starting point for conventional simulations~\cite{allen2017computer,frenkel2023understanding}, where the unphysical flips of the molecules are no longer proposed. This approach was followed in earlier studies employing swap MC~\cite{berthier2020how,Scalliet2022,Guiselin2022}. This is a useful strategy as equilibration is warranted, and several independent samples can easily be run in parallel to more efficiently perform an ensemble average. 

Several options are possible for the physical dynamics. First, we performed Monte Carlo simulations using only local translational moves of the atoms. These Monte Carlo simulations were also instrumental in benchmarking flip MC, as we could carefully check that energy distributions were identical in both types of Monte Carlo simulations. Second, we also used molecular dynamics (MD) simulations to study the physical relaxation of the model. As found for atomistic models, MD and local MC give essentially similar results~\cite{berthier2007monte}, \rev{in the sense that the long time decay of time correlation functions coincide up to a rescaling of the time. This property is useful as we can estimate the relative efficiency of Monte Carlo and Molecular Dynamics strategies to analyze the physical dynamics. We have found that the LAMMPS~\cite{thompson2022lammps} software for MD simulations is computationally more efficient than our MC dynamics implementation as it takes less CPU time to reach the same value of a given time correlation function. We therefore decided to report the results from MD simulations when it comes to the physical dynamics of the system.}

We perform MD simulations~\cite{allen2017computer} in the canonical $NVT$ ensemble using a time step $\delta t = 0.001$. The temperature is controlled using a Nosé-Hoover thermostat with a relaxation time set to $\tau = 1$. The initial velocities are assigned according to the Maxwell distribution at the desired temperature.

\begin{table}
\centering
\begin{tabular}{|ww|gg|ww|gg|}
\hline
$T$ & $\rho$ & $T$ & $\rho$ & $T$ & $\rho$ & $T$ & $\rho$ \\
\hline
2.93 & 1.00 & 2.03 & 1.12 & 1.73 & 1.14 & 1.21 & 1.18  \\
2.55 & 1.07 & 1.95 & 1.125 & 1.59 & 1.15 & 1.10 & 1.19\\
2.37 & 1.09 & 1.88 & 1.13  & 1.46 & 1.16 & 1.05 & 1.195 \\
2.19 & 1.11 & 1.80 & 1.135 & 1.33 & 1.17 & 1.00 & 1.20\\
\hline
\end{tabular}
\caption{The temperature and density path used to maintain a constant pressure $P=30\epsilon/\sigma_B^3$ in the $NPT$ ensemble.}
\label{tab:Trho}
\end{table}

In Sec.~\ref{sec:fragility}, we also perform simulations in the isothermal–isobaric $NPT$ ensemble. To identify equilibrium conditions at constant pressure $P$, we equilibrate samples in the $NVT$ ensemble with the flip MC algorithm over a range of temperatures and densities. This allows us to identify the thermodynamic path $\rho(T)$ for which the pressure remains constant. Subsequently, molecular dynamics simulations are performed in the $NVT$ ensemble along this path. In practice, we selected the value of the pressure $P$ as the pressure measured at low temperature $T = 1$ and density $\rho = 1.2$ in the $NVT$ ensemble, yielding $P= 30$. We then run simulations at temperatures and corresponding densities given in Table~\ref{tab:Trho}.

\section{Equilibration and algorithmic speedup}

\label{sec:equilibration}

\subsection{Equilibration runs and static quantities}

To use and demonstrate the capabilities of the flip Monte Carlo algorithm, we rely on the tests and experience developed earlier, when the swap Monte Carlo was applied at very low temperatures~\cite{ninarello2017models}. Briefly, we first validated the implementation of the flip MC algorithm by comparing the outcome with conventional MD results in the regime where both techniques can be used. We then used flip MC at decreasing temperatures, making sure that several observables (static and dynamic) converge to equilibrium and that simulations are long enough to accumulate sufficient statistics~\cite{ninarello2017models}. The measurement of time correlation functions then allows us to ensure, when needed, that statistically independent equilibrium configurations are produced. 

We start by characterizing the basic structural properties of the system using two complementary observables~\cite{barrat2003basic}. We measure the total atom-atom pair correlation function $g(r)$ 
\begin{equation}
g(r) = \frac{1}{4\pi N\rho r^2} \left\langle \sum_{i \neq j}^{N} \delta( r -  |{\bf r}_i - {\bf r}_j | ) \right\rangle ,
\label{eq:gr}
\end{equation}
and the corresponding  static structure factor 
\begin{equation}
S(q) = \frac{1}{N} \left\langle \sum_{i \neq j} e^{i {\bf q} \cdot ({\bf r}_i - {\bf r}_j)} \right\rangle. 
\label{eq:sq}
\end{equation}
In these definitions, ${\bf r}_i$ denotes the position of atom $i$, and the sum runs over all pairs of atoms. We also measured the corresponding quantities for the center of mass of the molecules (not shown), but they do not provide additional insight.

\begin{figure}
\includegraphics[width=\columnwidth]{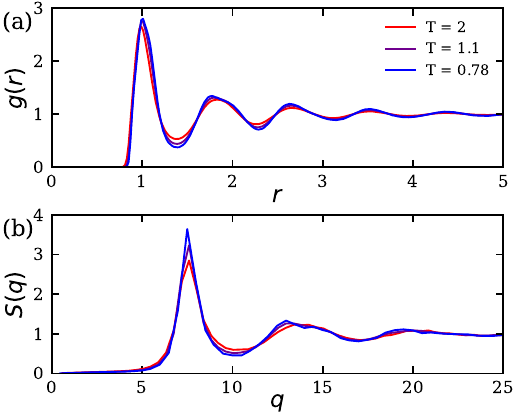}
\caption{Temperature evolution of two-point density correlations. (a) Radial distribution function of atoms. (b) Structure factor of atoms. No significant change in structure with temperature  is observed.}
\label{fig:structure}
\end{figure}

The results are shown in Fig.~\ref{fig:structure}, for three temperatures which cover a very broad range since the highest temperature is near the onset temperature, and the lowest temperature is close to $T_g$. The values of these characteristic temperatures are justified below. The measured two-point density correlators resemble observations made in atomic liquids, with a short-range structure that gradually becomes more pronounced when temperature decreases. This is consistent with the physical idea that in supercooled liquids, the structural evolution is very modest at the level of two-body correlation functions, with no signature of a phase transition. 

The unremarkable temperature evolution of the static structure in Fig.~\ref{fig:structure} demonstrates that our model, composed of an assembly of identical molecules with anisotropic shape, does not show any form of order as  temperature decreases. This was not guaranteed, as the temperature range explored here is unusually broad. Presumably, the very small amount of atomic diversity of the model compared to the Lewis-Wahnstr\"om model of ortho-terphenyl (which eventually crystallizes~\cite{pedersen2011crystallisation}) drastically enhances its glass-forming ability.  

\subsection{Time correlation functions} 

\label{sec:time}

To analyze the dynamic evolution of the system under flip Monte Carlo dynamics and to compare it with the physical MD dynamics, we examined a number of time correlation functions that characterize both translational and rotational motion. To define correlation functions that are not affected by any of the flip Monte Carlo moves, it is convenient to assign an index $j=1, \cdots, N$ to each particle. In a flip move, the type of the particle ($A$, $B$ or $C$) is exchanged but not the index, and we use indices to define and compute time correlations. This generalizes the strategy already used with swap MC~\cite{berthier2019efficient}.

The self-intermediate scattering function is defined as  
\begin{equation}
    F_s( q , t) = \left\langle \frac{1}{N} \sum_{j=1}^{N} e^{i \mathbf{q} \cdot (\mathbf{r_j}(t) - \mathbf{r_j}(0))} \right\rangle.
    \label{eq:fsqt}
\end{equation}
Here, $\mathbf{r}_j(t)$ represents the position of atom with index $j$ at time $t$. This position is unaffected by flip Monte Carlo moves. We average over wavevectors ${\rm q}$ with a modulus near $q = |\mathbf{q}| = 7.4$, corresponding to the first peak of the structure factor in Fig.~\ref{fig:structure}(b).

Similarly, an analogous self-intermediate scattering function can be defined to follow the motion of the center of mass of the molecules, namely 
\begin{equation}
    F_s^{\rm cm}(q, t) = \left\langle \frac{1}{M} \sum_{j=1}^{M} e^{i {\bf q} \cdot (\mathbf{r_j^{cm}}(t) - \mathbf{r_j^{cm}}(0))} \right\rangle.
\end{equation}
In this case, we choose wavevectors with an amplitude near $q=5.1$, determined from the structure factor associated with the centers of mass, and $\mathbf{r^{cm}}_j(t)$ denotes the position of the center of mass of molecule $j$ at time $t$. This position is unaffected by flip Monte Carlo moves. 

To characterize molecular orientational motion, we consider rotational correlators defined as  
\begin{equation}
    C_l(t) = \left\langle \frac{1}{M} \sum_{j=1}^{M} P_l \left( \mathbf{u_j}(t) \cdot \mathbf{u_j}(0) \right) \right\rangle,
    \label{eq:Cl}
\end{equation}
where $P_l(x)$ is the Legendre polynomial of order $l$, and $\mathbf{u}_j(t)$ represents the orientation vector of molecule $j$. The orientation vector $\mathbf{u}$ is defined at $t=0$ as the vector that connects the center of mass of molecule $j$ to the atom of type $B$ whose index is $k$.
(This choice is arbitrary and other directions could of course be studied.) At later times $t>0$ the vector $\mathbf{u}$ points toward the atom of index $k$, irrespective of its type. Thus, a flip move leaves unchanged the orientation vector. We focus on the case $l=2$, which is experimentally relevant. 

\begin{figure}
\includegraphics[width=\columnwidth]{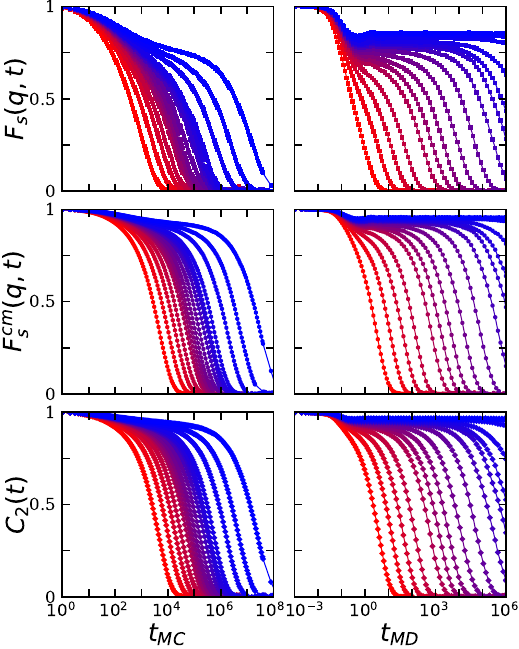}
\caption{Molecular rotational and translational correlation functions for flip MC dynamics (left column) and physical MD dynamics (right column). The temperatures are the same in both columns, from left to right: $T=3$, $2$, $1.6$, $1.4$, $1.25$, $1.2$, $1.15$, $1.1$, $1.05$, $1$, $0.97$, $0.95$, $0.92$, $0.85$, $0.8$, and $0.75$.}
\label{fig:relaxation}
\end{figure}

In the left column of Fig.~\ref{fig:relaxation}, we show the time dependence of these three correlation functions over a broad range of temperatures, from $T=3$ where dynamics is fast, down to $T=0.75$ which we estimate is slightly below $T_g$. These data are obtained by running flip MC simulations for 10 to 100 independent samples running for at least 100 times longer than the relaxation time at each temperature. The time decay of these functions to zero at all temperatures shows that translational motion of atoms and molecules slows down but is not arrested, and that the orientation of the molecules is also not frozen over the considered time window. As usual in glassy systems, the temperature evolution of time correlators is much more pronounced than static ones, and the dynamic relaxation of the system slows down as temperature decreases. At temperatures below $T=0.75$ it became difficult to ensure equilibrium and ergodic sampling of configuration space, although this should be possible using longer simulations.    

Using configurations equilibrated using the flip MC simulations we produce independent configurations from which we run conventional MD simulations, see Sec.~\ref{sec:MD}. The results for the MD dynamics are shown in the right column of Fig.~\ref{fig:relaxation}. For all three correlators, we observe a simple decay at high temperatures, followed by the development of a characteristic two-step decay of time correlations as $T$ decreases. When $T$ decreases further, a clear intermediate plateau appears, and the relaxation time grows rapidly. For the lowest temperatures studied, the correlation functions do not decay from the plateau over the accessible time window, even though the system is fully equilibrated. This observation shows that the relaxation dynamics using flip MC and using MD are controlled by vastly different timescales, and that flip MC more efficiently decorrelates particle configurations.  

\subsection{Relaxation timescales}

\label{sec:relaxation}

We define several structural relaxation times associated with translational and rotational motion. For translation, we define the relaxation time \( \tau_{\alpha} \) and the corresponding center-of-mass timescale \( \tau_{\alpha}^{\rm cm} \) from the time decay of the self-intermediate scattering functions  \( F_s(q, t) \), \( F_s^{\rm cm}(q, t) \), and  the orientational relaxation time \( \tau_{2} \) from that of \( C_2(t) \). The relaxation times are defined when these functions reach the value $e^{-1}$. 

\begin{figure}
\includegraphics[width=\columnwidth]{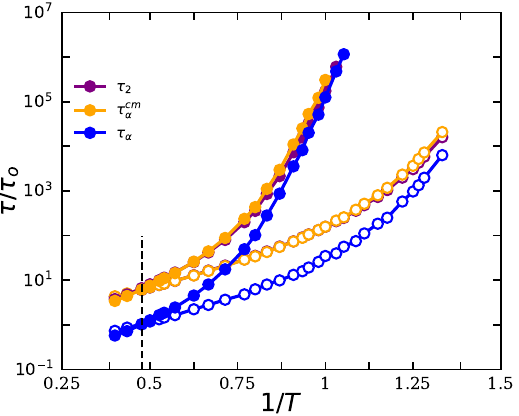}
\caption{Temperature evolution of three relaxation times defined from different time correlation functions for the physical MD dynamics (full symbols) and the flip MC dynamics (open symbols). Relaxation times are rescaled by the onset time $\tau_o$, independently defined for each dynamics. For physical dynamics, $\tau_o \approx 1.4$, and for the flip dynamics $\tau_o \approx 1.3 \times 10^3$. On this scale, $\tau_2$ and $\tau_\alpha^{\rm cm}$ are almost undistinguishable. The dashed segment indicates the onset temperature $T_o$.}
\label{fig:relaxation2}
\end{figure}

Figure~\ref{fig:relaxation2} shows relaxation times from three correlation functions and two types of dynamics (flip MC and MD). To enable comparison, all times are rescaled by an onset timescale $\tau_o$, defined by $\tau_\alpha(T_o) = \tau_o$, where $T_o \approx 2.1$ marks the onset of super-Arrhenius behavior in the MD dynamics. This yields $\tau_o \approx 1.4$ for MD and $\tau_o \approx 1.3 \times 10^3$ for flip MC.

For both dynamics we find that $\tau_2 \sim \tau_\alpha^{\rm cm}$ over the entire temperature range, showing that a significant rotation of the molecules takes place, on average, when the molecules also have moved significantly in space. Atomic translations are about ten times faster at high temperatures, but this difference decreases as $T$ decreases, so that $\tau_2 \sim \tau_\alpha^{\rm cm} \sim \tau_\alpha$ at very low temperatures and all timescales are strongly coupled.  

The most striking observation in Fig.~\ref{fig:relaxation2} is the very different temperature evolution observed for the two dynamics. Physical MD dynamics slows down considerably below $T_o$, and we observe a slowdown of about six orders of magnitude at $T = 0.95$. At this temperature, the flip MC has only slowed down by roughly two orders of magnitude, already demonstrating a speedup by a factor $10^4$ compared to MD. We can thus follow the flip MC dynamics in equilibrium conditions down to much lower temperatures, $T \approx 0.75$. Thus, we conclude that flip MC allows us to efficiently study the equilibrium physical properties of the molecular model at much lower temperatures than what was previously achievable.

\subsection{Equilibration speedup}

\begin{figure}
\includegraphics[width=\columnwidth]{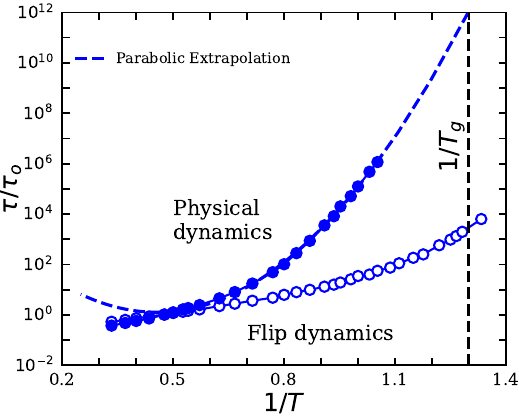}
\caption{Estimating the equilibration speedup of flip MC. The relaxation times $\tau_\alpha$ are rescaled by their respective onset timescale. The MD dynamics is extended to low $T$ using a parabolic fit, Eq.~(\ref{eq:parabolic}), from which the experimental glass transition temperature $T_g \approx 0.77$ is estimated. The equilibration speedup at the lowest simulated $T=0.75$ is about $3 \times 10^9$.}
\label{fig:acceleration}
\end{figure}

To provide a quantitative estimate of the speedup offered by the flip MC algorithm over the MD dynamics, we follow similar earlier attempts used in the context of the swap MC algorithm, see Fig.~\ref{fig:acceleration}. Using the relaxation timescale $\tau_\alpha$ rescaled by the respective onset timescales $\tau_o$, we fit the temperature evolution of the MD dynamics to a parabolic (in Arrhenius coordinates) expression~\cite{elmatad2010corresponding},
\begin{equation}
  \ln (\tau_\alpha/\tau_o ) = J^2\left( \frac{1}{T} - \frac{1}{T_p} \right)^2,
  \label{eq:parabolic}
\end{equation}
where $J \approx 6.2$ is the energy parameter and $T_p \approx 2.2$ represents an alternative definition for the onset temperature~\cite{keys2011excitations}. We observe that $T_p \simeq T_o$, as expected. The expression (\ref{eq:parabolic}) is favored as it is empirically found to extrapolate better to low temperatures~\cite{elmatad2010corresponding,ozawa2019does}.

We use the parabolic law to estimate the temperature evolution of $\tau_\alpha$ beyond the numerically accessible regime, see Fig.~\ref{fig:acceleration}. In particular, this allows us to provide a quantitative estimate of the experimental glass transition temperature $T_g$, which we define as $\tau_\alpha(T_g)/\tau_o = 10^{12}$~\cite{schmidtke2012from}. We obtain $T_g \approx 0.77$. At this temperature, the flip MC dynamics can still be simulated in equilibrium with an estimated speedup of about $10^9$ compared to the MD dynamics, see Fig.~\ref{fig:acceleration}. With a reasonable computational effort, the flip MC can provide equilibration and ergodic sampling down to $T=0.75$, which is below $T_g$. At this temperature, the estimated speedup is $3 \times 10^9$, compared to  molecular dynamics.  

The broad conclusion of this section is that the flip Monte Carlo algorithm provides, for a system composed of identical molecules, an equilibration speedup comparable to that of the swap Monte Carlo algorithm applied to multi-component and continuously polydisperse atomic systems, and thus reaches temperatures below $T_g$.  

\section{Signatures of glassy dynamics}

\label{sec:dynamics}

\subsection{Reaching the millisecond timescale}

Using the flip MC algorithm, we can efficiently prepare equilibrium configurations at arbitrarily low temperatures. \rev{We recall that the speedup offered by the flip MC dynamics stems from molecules performing unphysical moves (instantaneous rotations) that are constructed to sample more efficiently the configuration space at thermal equilibrium. To analyze the physical dynamics at these very low temperatures, one needs to revert to either Monte Carlo simulations with simple local moves, or to conventional molecular dynamics. As demonstrated in several earlier works~\cite{berthier2020how,Scalliet2022} the fast production of equilibrium configurations allows for more efficient dynamic studies, since one can launch conventional MD simulations from an ensemble of independent initial conditions. This strategy} allows us to explore the low-temperature  relaxation  dynamics of the model over a broader time window, as the entire MD trajectory can be used to collect useful data. In this section, we use this strategy to characterize the relaxation dynamics of the molecular model over a broad temperature range, allowing us to perform detailed comparisons with  numerical studies of atomistic models  or with experimental results on molecular fluids.

The discussion of fragility and decoupling phenomena is only possible because we can reach very long timescales. This very large time window is itself only possible because we can fully bypass the need to equilibrate the system using MD simulations, a step that is shortcut by the flip MC technique. Using the LAMMPS software and conventional CPU resources we can simulate about $10^9$ MD steps in about two weeks for $N=1000$ triatomic molecules. Given the small time discretization $dt = 10^{-3}$ needed to ensure numerical stability, we can then simulate up to $t = 10^6 \tau_o$, which, converted in experimental timescales (where $\tau_o \approx 10^{-10}$ps~\cite{schmidtke2012from}) yields a timescale of about 0.1ms. Using LAMMPS, we could of course easily use multiple CPUs and much longer simulations, as in Ref.~\cite{Scalliet2022} to reach simulation times reaching up to 1-10ms, if needed. 

%\textcolor{red}{run a long simu romain!}

\subsection{Fragility of molecular and atomistic models}

\label{sec:fragility}

The temperature evolution of the relaxation times in fragile glass-formers departs from an Arrhenius dependence. This observation led to the concept of glass fragility~\cite{angell1991,angell1995formation}, which quantifies the strength of these deviations. While several quantitative proposals of a definition of the glass fragility have been proposed, a relatively simpler comparison between systems can be used by starting from the following representation of the temperature evolution of the relaxation time:  
\begin{equation}
    \tau_{\alpha} = \tau_{\infty} \exp \left( \frac{E(T)}{T} \right) ,
\label{eq:energy}
\end{equation}
where $E(T)$ is the effective energy barrier, which depends on  temperature in fragile systems, and $\tau_{\infty}$ is a microscopic relaxation time. Our goal is to directly represent the temperature evolution of $E(T)$, in such a way that different systems can be compared. This is achieved by introducing $E_\infty$, the value of $E(T)$ at high temperature above the onset $T_o$ below which super-Arrhenius behavior is observed~\cite{tarjus2004disentangling,schmidtke2012from}. One can then compare the evolution of the rescaled quantity $Y = E(T)/E_\infty$ as a function of $X=T_o/T$ so that all systems fall on the curve $Y=1$ for $X \leq 1$. In this representation, a sharper dependence $Y(X)$ for $X>1$ reveals a larger glass fragility~\cite{KIVELSON199527}. 

\begin{figure}
\includegraphics[width=\columnwidth]{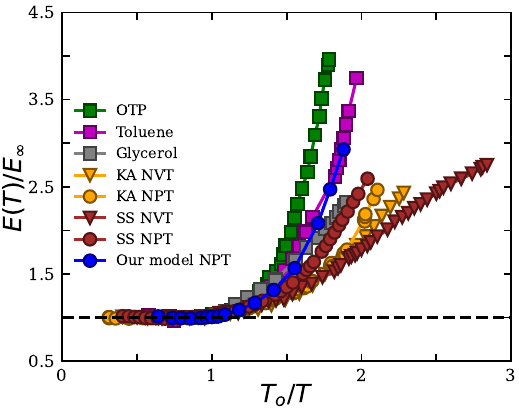}
\caption{\revv{Comparing the fragility of experimental and numerical glass-formers. The activation energy $E(T)$, Eq.~(\ref{eq:energy}), is rescaled by its high temperature value, $T \geq T_o$. \revv{Squares} represent experiments (data from Ref.~\cite{alba2022}), \revv{circles and triangles} are numerical models: Kob-Andersen (KA) isochoric data from Ref.~\cite{Coslovich2018} and isobaric from \cite{coslovich2007undrstanding}, polydisperse soft spheres in 3D: isochoric from Ref.~\cite{Scalliet2022} and isobaric from \cite{ozawa2019does}.}}
\label{fig:fragility}
\end{figure}

We construct this plot in Fig.~\ref{fig:fragility}. To ensure a relevant comparison with experiments, we use the results from simulations performed at constant pressure, as explained in Sec.~\ref{sec:MD}. As is well-known, glass fragilities are larger in isobaric conditions than in isochoric conditions, because the density of the system increases as  temperature decreases when pressure is imposed~\cite{tarjus2004disentangling}. The data in Fig.~\ref{fig:fragility} show that the fragility of our model is placed between that of glycerol and of ortho-terphenyl (OTP), and it is very close to that of toluene. 

To this plot we add two data sets obtained recently in two three-dimensional models of atomistic glass-formers. Using graphic cards, very low temperature data were obtained for the binary Kob-Andersen model~\cite{Coslovich2018}. The second data set is obtained by combining swap Monte Carlo with MD simulations to estimate the relaxation time of a three-dimensional soft sphere system with a continuous size polydispersity~\cite{Scalliet2022}. In both cases, it was noted that at very low temperatures, the evolution of the relaxation time was returning to a nearly Arrhenius regime. This is confirmed in the representation of Fig.~\ref{fig:fragility}, where the activation energy in the two models exhibit a tendency to saturate at very low temperatures, in a manner that is very different from experimental data. \revv{A faithful comparison between our model and previous studies using atomistic ones would require an extended dynamic range studies using an isobaric path. No such data set is available, but we nevertheless include data from two studies. As expected, these data confirm that fragility is slightly larger in such conditions, but the gap with molecules remains.}  

It has long been known that most numerical models of glass-formers are much less fragile than experimentally studied molecular fluids~\cite{tarjus2004disentangling}. Strikingly, the present model performs very well in this respect. A possible explanation is that molecules are characterized by more dynamic degrees of freedom (rotations and translations) compare to point particles, and that the simultaneous dynamic arrest of a larger number of strongly coupled degrees of freedom leads to a steeper temperature dependence. \revv{It would be interesting to more systematically test this idea using carefully designed molecular architectures and tailored intramolecular flexibility, as the latter is known to considerably influence the fragility of polymer glasses~\cite{dalle-ferrier2016}.} 

\subsection{Stokes-Einstein-Debye decoupling}

Since the seminal experimental work of Fujara {\it et al.}~\cite{fujara1992translational}, the breakdown of the Stokes-Einstein relations and related decoupling phenomena in supercooled liquids has spurred significant research efforts~\cite{cicerone1995molecules, cicerone1995relaxation, cicerone1996enhanced, mapes2006self, fujara1992translational, chang1997heterogeneity,sengupta2013breakdown}. It is commonly agreed that these observations are the direct consequence of the presence of dynamic heterogeneity, with different observables being dominated by different parts of a broad distribution of relaxation timescales~\cite{cicerone1996enhanced}. This hypothesis is captured by a variety of analytic models~\cite{tarjus1995breakdown,stillinger1994translation,Berthier_2005}, which are all built upon the idea that slow dynamics is not uniform in space and time. 

Most experiments compare the temperature evolution of the viscosity to the one of the self-diffusion constant of the molecules, $D_t$, as both are related by the Stokes-Einstein relation. The Stokes-Einstein-Debye relation instead compares $D_t$ to the rotational timescale $\tau_2$ defined from the time decay of $C_2(t)$, Eq.~(\ref{eq:Cl}). A decoupling of these two quantities thus leads to the idea that rotation and diffusion are decoupled~\cite{stillinger1994translation}. Finally, in atomistic models that lack orientational degrees of freedom, the latter relation is often replaced by a comparison between $\tau_{\alpha}$ and $D_t$~\cite{berthier2004time}, which relies on the largely confirmed hypothesis that in molecular systems, $\tau_\alpha \propto \tau_2$~\cite{kawasaki2017identifying}. 

\begin{figure}
\includegraphics[width=\columnwidth]{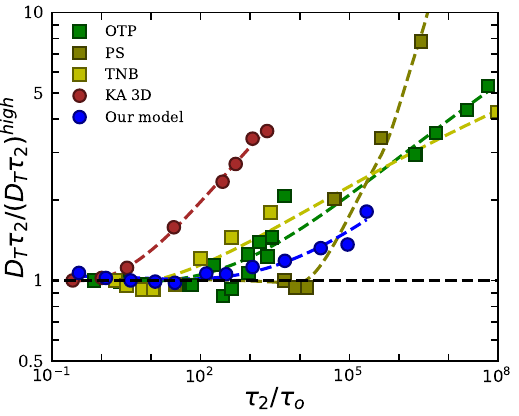}
\caption{Comparison of the Stokes-Einstein decoupling between our model, an atomistic model, and experiments. For atoms, we replace the $\tau_2$ with $\tau_\alpha$. The onset and amount of decoupling is comparable between our model and experiments, but the atomistic model shows an earlier and stronger decoupling. Squares represent experiments (OTP and trisnaphthylbenzene (TNB) data from \cite{mapes2006self} and polystyrene (PS) data from \cite{singlemoleculeDSE2022}), circles refer to numerical models (KA data from \cite{berthier2007monte}).}   
\label{fig:decoupling}
\end{figure}

To perform a quantitative comparison between our molecular model, atomistic simulations and experiments, we measured the temperature evolution of the product between the rotational correlation time $\tau_2$ and the self-diffusion coefficient $D_t$, which we normalize by its value in the high-temperature regime: $D_t \tau_2 / (D_t\tau_2)^{\text{high}}$. The diffusion constant $D_t$ is obtained in the long-time diffusive regime of the mean-squared displacement of the molecules. The results are shown in Fig.~\ref{fig:decoupling}. It is convenient to represent this product as a function of the relaxation time itself, $\tau_2(T)$, rescaled by its value $\tau_0 \sim 9$ at the onset temperature $T_o$. We thus expect  the data to be equal to unity at least when $\tau_2/\tau_0 < 1$.  Our simulations indicate that the Stokes-Einstein relation remains valid at temperatures lower than the onset $T_0$, and starts to be visible only for $\tau_2 / \tau_0 > 10^2$. 

A very different behavior is observed in atomistic models, with a strong decoupling developing already near the onset $T_o$, as illustrated in Fig.~\ref{fig:decoupling} for the well-studied Kob-Andersen binary mixture~\cite{berthier2007monte}, but also found across several numerical simulations. The quantitative difference is rather striking, because the amount of decoupling observed within the traditional time window of about four decades studied in conventional MD studies is actually nearly negligible in our molecular model. In fact, if atomistic models behaved like our molecular model, it would have been challenging to study decoupling phenomena using conventional MD methods before the development of the swap Monte Carlo algorithm.    

It is interesting to discuss this quantitative difference between atomistic and molecular models in the light of experiments. We add in Fig.~\ref{fig:decoupling} the decoupling data collected for three molecular fluids. They all demonstrate a delayed onset and a smaller amount of decoupling compared to atomistic models, with a very good agreement with our numerical findings. 

A possible interpretation is that the size dispersity, unavoidable to prevent the crystallization of spherical particles, introduces a source of structural heterogeneity that enhances the dynamic heterogeneity compared to molecular fluids~\cite{pihlajamaa2023influence}. This interpretation is supported by the measurement of the van Hove distribution function for the center-of-mass displacement of the molecules, where we observe weak deviations from non-Gaussianity at modest supercooling and only becomes non-Gaussian, with emerging exponential tails~\cite{chaudhuri2007universal}, at much lower temperatures and longer relaxation timescales compared to atomistic models. We conclude that molecular fluids display a dynamic heterogeneity qualitatively similar to size disperse atomistic models, but it is quantitatively less pronounced at comparable degree of slowing down, in better agreement with experiments.  

\section{Spatially heterogeneous dynamics}

\label{sec:dynamic}

In this section we examine whether the equilibrium relaxation of the system is spatially heterogeneous~\cite{berthier2011dynamical}. We demonstrate quantitatively the growth of dynamic heterogeneity as the dynamics slows down, characterizing both translational and orientational degrees of freedom. 

\subsection{Direct visualization of correlated domains}

\label{sec:direct}

We start with a real space illustration of the growth of spatial correlations in the dynamics. To this end, we resolve the relaxation at the particle scale, beyond the average correlators defined in Sec.~\ref{sec:time}. For each particle we define the contribution of atom $i$ to the self-intermediate scattering function, $F_s^i(t) = \cos ( {\bf q} \cdot \Delta {\bf r}_i(t))$, and the contribution of molecule $i$ to the orientational and translational correlators, $C_2^i(t) = P_2( {\bf u}_i(t) \cdot {\bf u}_i(0))$, and $F_s^{{\rm cm},i}(t) = \cos( {\bf q} \cdot \Delta{\bf r}_i(t)) $. 

\begin{figure}
\includegraphics[width=\columnwidth]{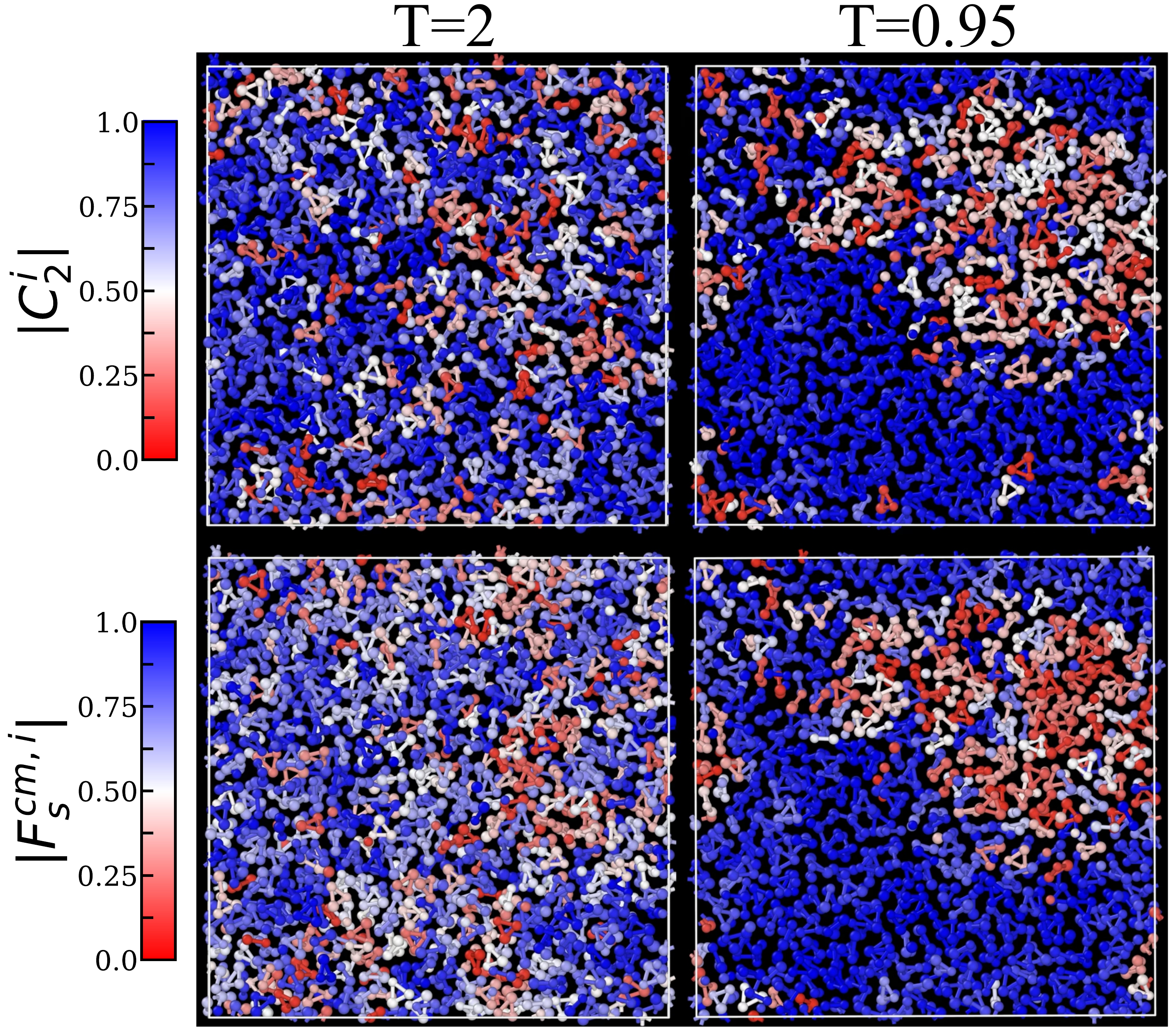}
\caption{Snapshots of slabs of size $3.0 \times 21.5 \times21.5$ taken from simulations with 4000 molecules, at two different temperatures and at times such that $C_2=0.65$ ($\tau_{\alpha}(T=2) / \tau_o \approx 1$ and $\tau_{\alpha}(T=0.95) / \tau_o \approx 10^6$). The top row shows orientational dynamic heterogeneities, and the bottom row translational ones, with color code evolving from immobile (blue) to very mobile (red). The emergence of spatial correlations and the strong correlation between orientational and translational motion are obvious.}
\label{fig:snapshot} 
\end{figure}

These definitions allow us to directly visualize the  fluctuations of the relaxation dynamics in real space, as demonstrated in Fig.~\ref{fig:snapshot}. We have chosen two temperatures, $T=2$ which is close to the onset temperature, and $T=0.95$ where the relaxation time has increased by a factor $10^6$ and the dynamics is very slow. In both cases, we choose an observation time such that $C_2(t) = 0.65$,  roughly in the middle of the structural relaxation. Finally we color code each particle according to the value of either $C_2^i(t)$ (top row) or $F_s^i(t)$ (bottom row), and we show a thin slab of the entire simulation box to better observe spatial correlations.  

Two key observations emerge from these images. Firstly, the very homogeneous fluctuation pattern observed at the onset is replaced at low temperature by large-scale correlations with obvious domains of fast moving molecules in red, immersed in large immobile regions. Therefore, the slowing down of the average dynamics is clearly accompanied by a growth of spatial correlation in the local dynamics. Secondly, the two correlated snapshots at low temperatures are nearly identical for the two observables, suggesting that molecules that rotate faster than the average, also translate more easily. In other words, dynamic heterogeneity in translational and orientational motion of the molecules appear to be strongly coupled in our model, consistent with the weak degree of decoupling between the corresponding timescales discussed in Sec.~\ref{sec:relaxation}. 

\subsection{Growing four-point susceptibilities}

\label{sec:four}

We now quantify these qualitative observations using appropriate statistical tools. The growth of spatial correlations of the dynamics can be quantified using four-point dynamic susceptibilities. As we defined several correlators, several susceptibilities can also be defined. For a given ensemble-averaged time correlation function, $C(t) = \langle 1/N \sum_i C_i(t) \rangle = \langle \hat{C}(t) \rangle$, one can define its instantaneous value $\hat{C}(t)$, and the instantaneous contribution of particle $i$, $C_i(t)$. The four-point susceptibility $\chi_4(t)$ is defined as the variance of the spontaneous fluctuations of $\hat{C}(t)$~\cite{berthier2011dynamical}:
\begin{equation}
\chi_4 (t) = M (\langle \hat{C}^2(t) \rangle - \langle \hat{C}(t) \rangle^2),
\label{eq:chi4}
\end{equation}
where the prefactor is $M$ for a molecular correlation (such as $C_2$ and $F_s^{\rm cm}(t)$, or $N$ for a correlation involving single atoms (such as $F_s(t)$).

\begin{figure}
\includegraphics[width=\columnwidth]{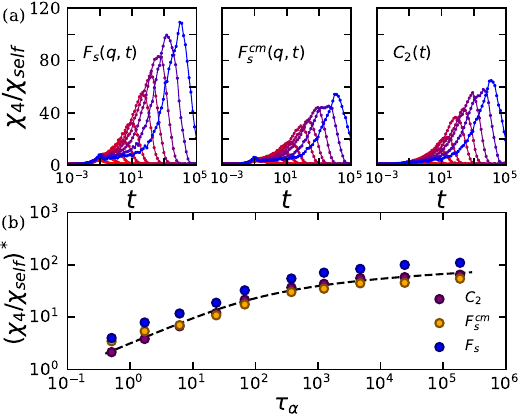}
\caption{(a) Time dependent four-point susceptibilities normalized by their self values, for $T=3$, $2$, $1.6$, $1.4$, $1.3$, $1.2$, $1.15$, $1.1$, $1.05$, and $1$ (from left to right), and three different observables. (b) Parametric evolution of the peak values of the normalized $\chi_4$ against the relaxation time $\tau_\alpha$. The dashed line represents a fit to the $C_2$ data using Eq.~(\ref{eq:dalle-ferrier}).}
\label{fig:chi4} 
\end{figure}

The first square in the right hand side of Eq.~(\ref{eq:chi4}) contains all the relevant information about the correlations between different molecules, as it is a sum of terms of the form $\langle C_i(t) C_j(t) \rangle$. It is straightforward to show that in the absence of any intermolecular \rev{correlations} the only contribution to $\chi_4(t)$ is the self-part, defined as:   
\begin{equation}
\chi_4^{\rm self}(t) = \langle C_i^2 (t) \rangle - \langle C_i (t) \rangle^2,
\end{equation}
which only contains information about single-particle fluctuations. It is therefore convenient to rescale $\chi_4$ by $\chi_4^{\rm self}$ so that by definition the ratio is unity in the absence of any correlations between molecules, whereas a value larger than unity indicates a growing amount of correlations, which provides a good quantitative estimate of the volume of the spatial correlations~\cite{lacevic2003spatially,toninelli2005dynamical,karmakar2014growing}.  

The time evolution of the rescaled four-point susceptibilities measured from the fluctuations of $F_s(q,t)$, $F_s^{\rm CM}(q,t)$, and $C_2(t)$ are displayed in Fig.~\ref{fig:chi4}(a) for a broad range of temperatures. By construction, the rescaled functions are close to unity at both short and long times, when correlations vanish, and are maximal for a time $t$ close to the average relaxation time of the respective average functions at each temperature. The three susceptibilities display very similar behavior, but reach a peak amplitude that depends weakly on the studied function, as discussed below.  

In Fig.~\ref{fig:chi4}(b), we report the temperature evolution of the peak of each rescaled susceptibility when temperature decreases. We follow previous work and replace temperature by the corresponding value of the relaxation time. We arbitrarily choose $\tau_\alpha(T)$ defined from $F_s(q,t)$ in Eq.~(\ref{eq:fsqt}) for the horizontal axis. This representation is useful, as it allows to directly assess the interrelation between growing timescales and growing lengthscales.  In this log-log representation, we confirm the rapid growth of correlation volumes for all three observables when $\tau_\alpha$ increases. The initial growth can be roughly described by a power law, $\chi_4 \propto t^{1/2}$, but this evolution becomes much more modest at lower temperatures~\cite{berthier2007spontaneous,Coslovich2018,das2022crossover}. 

We fit the peak values using the empirical functional form introduced in Ref.~\cite{dalle2007spatial}:
\begin{equation}
\tau_{\alpha} \simeq A\left[\left(\frac{\chi_4}{\chi_{\text{self}}}\right)^*\right]^{\gamma} \exp\left\{ \left[ B\left(\frac{\chi_4}{\chi_{\text{self}}}\right)^* \right]^{\psi} \right\},
\label{eq:dalle-ferrier}
\end{equation}
%previous values where the fitting parameters are \( A = 0.08 \), \( B = 1.3 \times 10^{-5} \), \( \gamma = 2 \), and \( \psi = 3 \). 
where the fitting parameters are $A = 0.10$, $B = 1.22 \times 10^{-3}$, $\gamma = 2$, and $\psi = 2$. 
As discussed in \cite{dalle2007spatial}, these parameter values should be considered indicative rather than definitive, as a range of alternative parameter sets can yield similarly good fits to the data. Nevertheless, our fitted values are in close agreement with those reported in the literature.
We also observe that the values of $\chi_4$ for $F_s^{\rm cm}$ and $C_2$ are nearly identical, suggesting again that orientational and translational degrees of freedom are strongly coupled and display similar spatial correlations, as hinted by the snapshots in Fig.~\ref{fig:snapshot}. In a similar vein, the fluctuations of $F_s(q,t)$ appear larger than for $F_s^{\rm cm}$ but with an identical temperature evolution. This is rationalized by the fact that the number of correlated atoms should be three times larger than the number of correlated molecules for fully rigid molecules. In practice the ratio of susceptibility is compatible with a slightly smaller factor of about 2.      

\subsection{Rotation-translation coupling at molecular scale}

\begin{figure}
\includegraphics[width=\columnwidth]{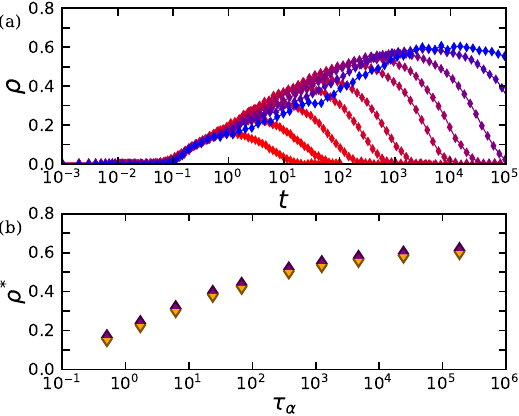}
\caption{(a) Time dependence of the Pearson correlation between individual molecular rotations and translations, Eq.~(\ref{eq:pearson}) for the same temperatures as in Fig.~\ref{fig:chi4}.
(b) Parametric evolution of the peak value of the Pearson correlation with the relaxation time $\tau_\alpha$. The correlation between individual translations and rotations grows rapidly to reach $\rho \approx 0.6$ at the lowest temperature.}
\label{fig:pearson} 
\end{figure}

We now take a closer look at the strong coupling between translations and rotations of the molecules visualized in Fig.~\ref{fig:snapshot}. To quantify whether the correlation holds at the molecular scale, we define the Pearson correlation coefficients between local fluctuations of $C_2$ and $F_s^{\rm}$,
\begin{equation}
\rho (t)=\frac{\langle C_2^i F_s^{{\rm cm},i} \rangle - C_2 F_s^{\rm cm}} 
{\sqrt{{\rm var}(C_2^i) {\rm var}(F_s^{{\rm cm},i})}} ,
\label{eq:pearson}
\end{equation}
with $C_2^i(t)$ and $F_s^{{\rm cm} , i}(t)$ the instantaneous values for molecule $i$ defined in Sec.~\ref{sec:direct}, and ${\rm var}(x)$ the variance of variable $x$. 

The time evolution of the Pearson coefficient is shown in Fig.~\ref{fig:pearson}(a). It vanishes at both short and long times, and exhibits a maximum near the relaxation time $\tau_\alpha$. This behavior resembles the one of the four-point functions in Fig.~\ref{fig:chi4}. The evolution of the value $\rho^*$ of the peak, which indicates the maximum of the cross-correlations, is reported in Fig.~\ref{fig:pearson}(b). As the relaxation time increases, $\rho^*$ increases quite rapidly, and reaches a strong value of about $\rho^* \approx 0.6$ at the lowest simulated temperature. Therefore, even at the level of individual molecules, correlations between local fluctuations of rotational and translational dynamics are quite strong. Similar observations were obtained for the simulated dynamics of supercooled water molecules~\cite{mazza2006relation}.

\section{Emergence of excess wings}

\label{sec:excess}

\subsection{Dynamical susceptibility}

We now turn to the analysis of dynamic relaxation spectra in Fourier space, as these measurements are routinely performed experimentally. In experiments, $\chi{''}(\omega)$ can be obtained from dynamic depolarized light scattering experiments or from broadband dielectric spectroscopy. These two techniques are respectively linked by a Fourier transform to the collective rotational correlators $C_2^{\rm tot}(t)$ and $C_1^{\rm tot}(t)$~\cite{Gabriel2018, Petzold2013}. Since the amplitude of the cross-terms in $C_2^{\rm tot}(t)$ is small, we can safely analyze dynamic spectra by considering the Fourier transform of $C_2(t)$ defined in Eq.~(\ref{eq:Cl}), which only contains the self-contributions~\cite{henot2023orientational}. 

Following earlier work, we calculate the dynamical susceptibility starting from the distribution of relaxation times $G(\log(\tau))$ which we obtain through the relation $G(\log(\tau)) \simeq dC_2(\tau) / d\log(\tau)$~\cite{Blochowicz2003,berthier2005numerical}, to obtain
\begin{equation}
\chi{''}(\omega)=\int^{+\infty}_{-\infty}G(\log(\tau))\frac{\omega \tau}{1+(\omega \tau)^2}d\log(\tau).
\end{equation}

\begin{figure}
\includegraphics[width=\columnwidth]{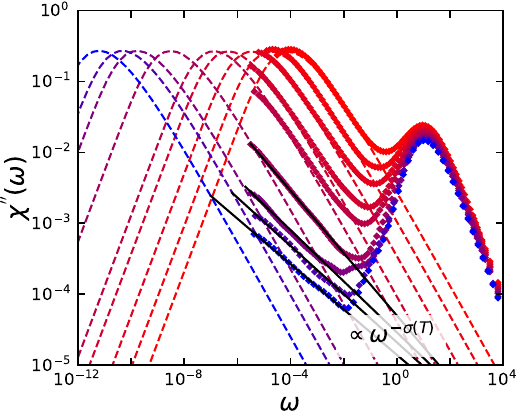}
\caption{Dynamic relaxation spectra for $T=1.1$, $1.05$, $1$, $0.95$, $0.92$, $0.85$, $0.8$, $0.78$, $0.75$, from the Fourier transform of $C_2(t)$. The dashed lines represent the structural relaxation peaks, either directly measured or estimated. At very low temperatures, an excess wing appears described by a power law decay (full lines) with a $T$-dependent exponent $\sigma(T)$.}
\label{fig:wings}
\end{figure}

The results are displayed in Fig.~\ref{fig:wings}. The dynamical susceptibility $\chi{''}$  of glass-formers presents two main peaks, which mirrors the two-step decay of correlation functions in the time domain. The high-frequency peak corresponds to rapid vibrations of the molecules. It occurs near $\omega \sim 10$ and its position varies very weakly with temperature. 

The second peak occurs at lower frequencies and corresponds to the structural (or $\alpha$-) relaxation. It rapidly shifts to lower frequencies as temperature decreases. At very low temperatures, obtaining the relaxation spectra becomes statistically very demanding. To obtain smoother data, we use the spline method described in Ref.~\cite{Guiselin2022}, which was shown to preserve the accuracy of the results. Also, near the experimental glass transition temperature $T_g$, we cannot run long enough simulations to directly observe the peak corresponding to the structural relaxation. Instead, we estimate their shape and location by carefully tracking the temperature evolution of $C_2(t)$ at higher temperatures. These data are fitted, at long times, with a stretched exponential decay, $C_2(t) \sim \exp(-(t/\tau_2)^{\beta})$, with $\beta$ the stretching exponent. \rev{Over a broad temperature range well below the onset $T_o$, we observe that a single value $\beta \approx 0.67$ describes our data very well ($\beta$ increases weakly to reach $\approx 0.8$ near the onset). This observation implies that time-temperature superposition is obeyed to a good approximation for this model, a property which has been discussed extensively in experiments~\cite{niss2018searching}. We also note that the observed stretching exponent $\beta =0.67$ is significantly distinct  from the  value $\beta=0.5$ often reported in molecular liquids~\cite{bohmer2025on}.} The peak locations are estimated by extrapolating $\tau_2(T)$ with an Arrhenius law. This approach underestimates the actual values of $\tau_2(T)$ and places the peak positions at frequencies that are overestimated. As a result, the excess contribution identified in the relaxation spectra are, in general, underestimated. Despite this conservative extrapolation, Fig.~\ref{fig:wings}, presents reasonable estimates of the structural relaxation peaks including those near $T_g$.

The representation of the low-frequency peaks is useful, as it reveals that the measured spectra at frequencies intermediate between the microscopic and the structural relaxation times are considerably larger than the sum of these two peaks. There is therefore a signal  in excess, and this ``excess wing" takes the form of a power law, $\chi{''} \propto \omega^{-\sigma(T)}$ with $\sigma(T)$ an exponent that decreases with decreasing temperature. These power law fits are shown in Fig.~\ref{fig:wings}, with exponents $\sigma$ decreasing from $0.44$ to $0.32$ at the lowest temperature. These very low values are not compatible with the extrapolation of the high-frequency flank of the $\alpha$ relaxation (which would scale as $\omega^{-\beta}$).   

The characteristics of the excess contribution observed in the dynamic spectra are quantitatively analogous to the excess wings reported experimentally~\cite{dixon1990scaling,leheny1998dielectric,menon1995evidence,schneider2000excess}, and observed numerically only much more recently~\cite{Guiselin2022}. In contrast to these recent atomistic simulations, our data are obtained for a molecular glass-former (instead of a continuously polydisperse atomic model), and using an orientational observable (instead of a translational correlator). Our results therefore strengthen the interpretation that excess wings can be observed in numerical simulations, with characteristics that very close to experimental results. In particular, the notable temperature evolution reported in Fig.~\ref{fig:wings} was not visible in earlier simulations, but it is in good agreement with experiments~\cite{dixon1990scaling}.  

\subsection{Microscopic interpretation of the wings}

The excess wings in the dynamic spectra have two major features. In absolute amplitude, the corresponding signal is about a hundred times smaller than the main peak corresponding to structural relaxation. The second feature is the power law behavior, $\chi{''} \sim \omega^{-\sigma(T)}$. 

\begin{figure}
\includegraphics[width=\columnwidth]{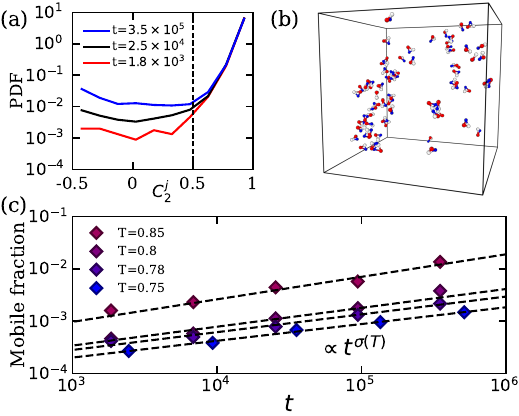}
\caption{(a) Probability distribution function of $P_2(\cos \theta_j(t))$ at $T=0.85$ at three different times. We define mobile particles using  the threshold shown as a vertical dashed line, in the tail of the distributions. (b) Snapshot showing only mobile molecules at $T=0.85$ and $t=10^6$. (c) Time evolution of the population of mobile particles at temperatures where an excess wing is visible in dynamic spectra. The growth is well described by a power law with the same exponent $\sigma(T)$ as in the excess wings in Fig.~\ref{fig:wings}.}
\label{fig:interpretation}
\end{figure}

Since we have access to the motion of each molecule in the time regime corresponding to the excess wings, we analyze the shape of the probability distribution of angular displacements, $P_2(\cos \theta_i(t))$, where $\theta_i(t)$ is the angular displacement of molecule $i$ over a time delay $t$. The data shown in Fig.~\ref{fig:interpretation}(a) confirm that after the time delay corresponding to the excess wings, a large majority of the molecules have rotated by a very small angle and display values of $P_2 (\cos \theta(t))$ close to unity. However, there is a small population of molecules characterized by much larger rotations and much lower values of $P_2$. 

In real space, these molecules correspond to rare locations in the system where small clusters of molecules perform large rotations much before the majority of the molecules, see Fig.~\ref{fig:interpretation}(b) for a representative snapshot. 

The connection between these localized clusters of molecules relaxing much before the bulk and the excess wings can be made quantitative. From the distributions in Fig.~\ref{fig:interpretation}(a), we define a threshold $P_2 = 0.5$ to distinguish between mobile ($P_2 < 0.5$) and immobile ($P_2>0.5$) molecules. We can then follow, for each temperature, the time evolution of the population of mobile particles. We perform these measurements for the four lowest temperatures where the excess wings are most pronounced, and we also focus on the time domain corresponding to the wings. The results are shown in Fig.~\ref{fig:interpretation}(c). Remarkably, we observe that the population of mobile particles defined above increases as a power law of time, $t^{\sigma(T)}$, with the same exponent $\sigma(T)$ that quantifies the excess wings in the spectra. The growth is therefore extremely slow (the exponents are in the range $\sigma \approx 0.32-0.44$) and it becomes slower at lower temperatures. These power laws presumably stem from a broad distribution of relaxation time with a power law tail, as observed in atomistic systems before~\cite{Guiselin2022}. 

This analysis demonstrates that a small number of particles that rotate significantly are at the origin of the excess wing. They generalize and reinforce the findings obtained for atomistic models using the swap Monte Carlo algorithm~\cite{Guiselin2022}. 

\section{Discussion and outlook}

\label{sec:discussion}

We proposed and implemented a computational strategy to devise simple molecular models for glass-formers for which efficient Monte Carlo moves can significantly speedup the equilibration time when the glass transition is approached. \rev{As usual with  accelerated Monte Carlo approaches, this strategy efficiently produces an ensemble of equilibrium configurations down to very low temperatures, but does not correspond to a physical dynamics of the system. However, the resulting configurations can subsequently  be used as initial conditions for conventional Molecular Dynamics simulations, in order to also analyze more efficiently the physical relaxation dynamics of supercooled liquids at thermal equilibrium~\cite{berthier2020how,Scalliet2022}.} 

As a proof of concept, we constructed a coarse-grained model of a molecule with a triangular geometry inspired by ortho-terphenyl, a well-studied molecular glass-forming liquid. We developed a flip Monte Carlo algorithm which we used to efficiently equilibrate and sample configurations at temperatures that are inaccessible to conventional molecular dynamics approaches, with an estimated speedup in equilibration timescales approaching $10^{10}$ at the lowest temperature. Our main contribution is thus to make the comparable performance of the swap Monte Carlo algorithm devised for atomic liquids available to molecular fluids. We thus anticipate that the explosion of numerical studies afforded by swap MC~\cite{berthier2023modern} will now extend to the world of molecules, thus closing an important gap between numerical and experimental studies of the glass transition.     

Our study is the first application of this general strategy. Yet, it already yields valuable insights into key physical questions concerning the glass transition in molecular systems. In particular, we demonstrated a significant increase in glass fragility compared to most atomistic models. We believe that our strategy paves the way towards a microscopic, molecular understanding of this most discussed property characterizing glassy liquids~\cite{alba2022}.   

We also demonstrated that a detailed analysis of molecular motion becomes possible at temperatures approaching the experimental glass transition. This shows that computer simulations can now directly address the question ``How do molecules move near $T_g$?'' posed thirty years ago by Cicerone {\it et al.}~\cite{cicerone1995molecules}. For the present molecular system, we observed a strong dynamic coupling, at the molecular scale, between rotational and translational dynamics. This observation has deep physical impact on several aspects, such as Stokes-Einstein decoupling, spatially heterogeneous dynamics, and the emergence of excess wings in relaxation spectra. In future work, we will in particular quantify similarities and differences with atomistic models, and revisit the notion of rotation-translation decoupling invoked in observed from the Stokes-Einstein relation. Another important line of research concerns the relevance of dynamic facilitation~\cite{chandler2010dynamics} (which is quite strong in atomistic models~\cite{Scalliet2022,herrero2024direct}) in the relaxation dynamics of molecular fluids and the role it potentially plays in controlling the asymmetric shape of dynamic relaxation spectra~\cite{Guiselin2022,scalliet2021excess}.  

Several other lines of research are enabled by our work, in order to address important physics question regarding glassy molecular systems. For instance, we can now prepare ultrastable molecular glasses~\cite{Ediger2017highly} and compare their properties to experimental systems, for instance to study the interplay between glass anisotropy and glass stability~\cite{bagchi2020controlling}. We can also analyze mechanical properties of molecular glasses prepared in conditions that are much closer to experimental systems~\cite{Ozawa2018}, as well as low-temperature excitations (such as harmonic density of states~\cite{wang2019low} and two-level tunneling systems~\cite{khomenko2020depletion}). It would also be interesting to revisit the thermodynamic properties of the supercooled liquid~\cite{Berthier2017}, to possibly confirm the recent experimental suggestion~\cite{beasley2019vapor} that molecular liquids still exist a few Kelvins above the Kauzmann temperature, much below $T_g$.   

On a different front, another avenue made possible by our work involves the extension of the flip Monte Carlo strategy, specifically developed here for the triangular geometry of the studied molecule, to a broader range of molecular geometries. With hindsight, it is clear that intramolecular Monte Carlo moves that approximately preserve the symmetry and the chemistry of the molecule, as flip MC does for the triangular shape shown in Fig.~\ref{fig:model}, can be devised for \rev{several other geometries, such as squares, pyramids, or tetrahedra, to name just the simplest ones. We are currently developing models for copolymer glasses~\cite{roth2016polymer} for which equilibration is notoriously difficult and where coarse-grained simulations have been frequently used~\cite{grest1986molecular,barrat2010molecular}. We thus believe that the approach proposed in this manuscript will be instrumental to tackle a broad variety of physics problems.} In future work we plan to systematically explore and demonstrate these possibilities in order to construct an extended zoo of molecular models where Monte Carlo algorithms generalizing the swap and flip Monte Carlo algorithms become readily accessible to a broader community of researchers interested in molecular glass-formers. \rev{This ambitious strategy will then open the possibility to disentangle universal from non-universal aspects of glass physics by the systematic observation of how molecular geometry, architecture, or intramolecular degrees of freedom impact many glassy properties, from the (possibly universal~\cite{bohmer2025on}) shape of dynamic correlation functions to the elucidation of the factors that influence glass fragility.} 

\section*{Data Availability}

The data that support the findings of this study are openly available on Zenodo~\cite{SimonZenodo2026}.

\acknowledgments  

We thank C. Alba-Simionesco, M. Ediger, M. H\'enot, C. Herrero, F. Ladieu, C. Scalliet, and G. Tarjus for many interesting discussions. L.B. acknowledges the support of the French Agence Nationale de la Recherche (ANR), under grants ANR-20-CE30-0031 (project THEMA) and ANR-24-CE30-0442 (project GLASSGO).

\bibliography{sample.bib}

\end{document}